**CERN for AI: A Theoretical Framework for Autonomous Simulation-Based Artificial Intelligence Testing and Alignment**

Ljubiša Bojić[1], Ph. D., Senior Research Fellow
The Institute for Artificial Intelligence Research and Development of Serbia, Novi Sad, Serbia
University of Belgrade, Institute for Philosophy and Social Theory, Digital Society Lab, Belgrade, Serbia

Matteo Cinelli[2], Ph. D., Assistant Professor
University of Rome La Sapienza, Department of Social Sciences and Economics, Rome, Italy

Dubravko Ćulibrk[3], Ph. D., Full Professor
The Institute for Artificial Intelligence Research and Development of Serbia, Novi Sad, Serbia

Boris Delibašić[4], Ph. D., Full Professor
University of Belgrade, Faculty of Organizational Sciences, Belgrade, Serbia

**Abstract** This paper explores the potential of a multidisciplinary approach to testing and aligning artificial intelligence (AI), specifically focusing on large language models (LLMs). Due to the rapid development and wide application of LLMs, challenges such as ethical alignment, controllability, and predictability of these models emerged as global risks. This study investigates an innovative simulation-based multi-agent system within a virtual reality framework that replicates the real-world environment. The framework is populated by automated 'digital citizens,' simulating complex social structures and interactions to examine and optimize AI. Application of various theories from the fields of sociology, social psychology, computer science, physics, biology, and economics demonstrates the possibility of a more human-aligned and socially responsible AI. The purpose of such a digital environment is to provide a dynamic platform where advanced AI agents can interact and make independent decisions, thereby mimicking realistic scenarios. The actors in this digital city, operated by the LLMs, serve as the primary agents, exhibiting high degrees of autonomy. While this approach shows immense potential, there are notable challenges and limitations, most significantly the unpredictable nature of real-world social dynamics. This research endeavors to contribute to the development and refinement of AI, emphasizing the integration of social, ethical, and theoretical dimensions for future research.

[1] Corresponding author;
Email address: ljubisa.bojic@ivi.ac.rs
Address of correspondence: Fruskogorska 1, 21000 Novi Sad, Serbia;
ORCID: 0000-0002-5371-7975
[2] Email address: matteo.cinelli@uniroma1.it
Address of correspondence: Piazzale Aldo Moro 5, 00185 Rome, Italy;
ORCID: 0000-0003-3899-4592
[3] Email address: dubravko.culibrk@ivi.ac.rs
Address of correspondence: Fruskogorska 1, 21000 Novi Sad, Serbia;
ORCID: 0000-0003-3417-1687
[4] Email address: boris.delibasic@fon.bg.ac.rs
Address of correspondence: Jove Ilica 154, 11000 Belgrade, Serbia;
ORCID: 0000-0002-6153-5119

## Introduction

The rapid evolution and expansion of artificial intelligence (AI), especially in the domain of natural language processing (NLP), has proven to be a promising frontier in technological development. AI-driven applications, particularly those based on Generative Pretrained Transformers (GPT), possess the potential to revolutionize various sectors of society by transforming processes, interactions and services, presenting many possibilities that were previously unimaginable (Chen et al., 2020). The burst of innovative technologies based on AI, such as recommendation algorithms, chatbots, autonomous vehicles, and even complex financial trading strategies, have somewhat become part of our daily lives, integrating their functionalities globally across numerous industries and sectors.

With the increased reliance and adoption of such AI systems, numerous challenges pertaining to the alignment with human ethics and values, controllability, transparency and predictability of these models arise, therefore warranting further attention and investment in terms of research and development. While AI has made significant strides in decision-making and task-completion competencies, achieving alignment with human values, predictability and full controllability, especially for large-scale neural networks, remains a stumbling block in the evolution of this powerful technology (Leike et al., 2017).

Recent contributions in AI research have focused on Generative Pretrained Transformers (GPT), a model that employs machine learning algorithms to improve the generation of human-like text (Radford et al., 2019). However, the rise of these highly-potent AI models has amplified concerns about their capacity for ethical alignment, controllability, and the unpredictability often inherent in large-scale neural networks (Irving & Askell, 2019).

There is insufficient understanding of how these models would behave in complex social dynamics and unfolding scenarios mirroring the real-world. This has put pressure on AI stakeholders to explore better testing and mitigation strategies (Chen et al., 2020). Simultaneously, the need to ensure AI models' security is an imperative concern which is made paramount due to the profound potential impacts of implementing them for societal and commercial purposes (Chen et al., 2020; Irving & Askell, 2019).

### *Definitions*

Large language models (LLMs) are artificial intelligence neural networks capable of language generation, translation, question answering and summarization (Radford et al., 2019). Aside from text generation, LLMs also exhibit the capacity to simulate understanding of inquiries and perform complex cognitive tasks (Sartori & Orrù, 2023). LLMs have demonstrated relatively high performance in a wide range of tasks and languages without any task-specific training (Radford et al., 2019).

Artificial General Intelligence (AGI) is a futuristic concept, which refers to a type of AI with cognitive capabilities that can successfully understand, learn, and implement wide range of intellectual tasks equivalent to those of a human being (Gartner, 2023).

AI Alignment represents the proposition of ensuring that the behavior of AI system is congruent with human intentions and values. The development of AI might lead to an intelligence explosion where AI surpasses human intelligence. If such a situation arises, it is important to ensure that AI is beneficially aligned and promotes the interests of humanity (Bostrom, 2014). Thus, research is needed to ensure that AI development is carried out with necessary precautions.

Among other platforms, OpenAI's LLMs stand out due to their potential for fine-tuning, making them compatible with a wide range of use-cases. This adaptability sets the stage for their comprehensive influence and application across diverse fields. OpenAI develops AGI while attempting to devise strategies that ensure its safety and alignment with human values (Altman, 2023).

*Practical applications leading towards simulation-based AI testing*

LLMs can be given various degrees of autonomy while creating multiple agents with different prompts capable of interacting with each other. There are three notable applications of LLMs in the direction of simulations and autonomy: Auto-GPT (Lutkevich, 2023), Interactive LLM Powered NPC (AkshitIreddy, 2023) and AI Town (Convex, 2023).

Auto-GPT is an experimental, open-source autonomous AI agent, created on the underlying principles of the GPT-4 language model (Lutkevich, 2023). It is designed to autonomously chain together various tasks in order to achieve a bigger, overarching goal as set by the user. Unlike traditional chatbots like ChatGPT that require multiple prompts to function, Auto-GPT operates by automating this multi-step prompting procedure. The user merely has to provide a single initial prompt or a set of instructions coded in natural language, and Auto-GPT handles the rest, breaking down the provided goal into a series of manageable subtasks to accomplish its objective. Thus, Auto-GPT can be employed in similar ways as ChatGPT, but with the added advantage of automation, thereby ensuring quicker task completion. It also boasts internet integration, thereby allowing it to access and use real-time data. Logo of Auto-GPT is depicted on Figure 1.

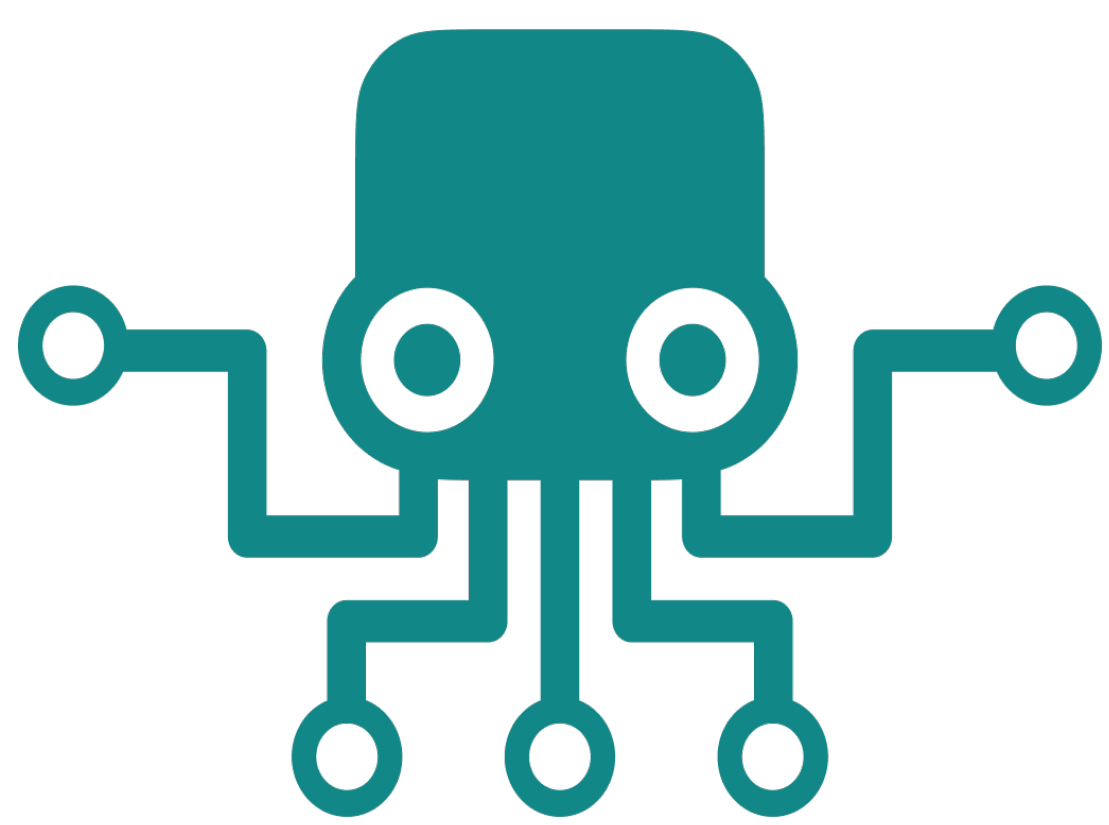

Figure 1. A simple digital vector art of an octopus like creature, used as the logo of Auto GPT (AutoGPT, 2023).

Interactive LLM Powered NPCs is an open-source project aiming to improve player interaction with non-player characters (NPCs) in video games (AkshitIreddy, 2023). The project transforms static conversations with NPCs to dynamic ones, allowing players to engage in immersive dialogues, recognizing their voice and showing lifelike animations of the characters (Figure 2). The technology used includes facial animation to sync character lip movements, facial recognition to differentiate characters, and vector stores to give NPCs limitless memory capacity. It also uses a pre-conversation file to shape the dialogue style of each character, making the interaction more lifelike. The NPC adjusts based on its specific personality, knowledge, and communication style, and is capable of perceiving the player's facial expressions via webcam, adding depth to the interaction.

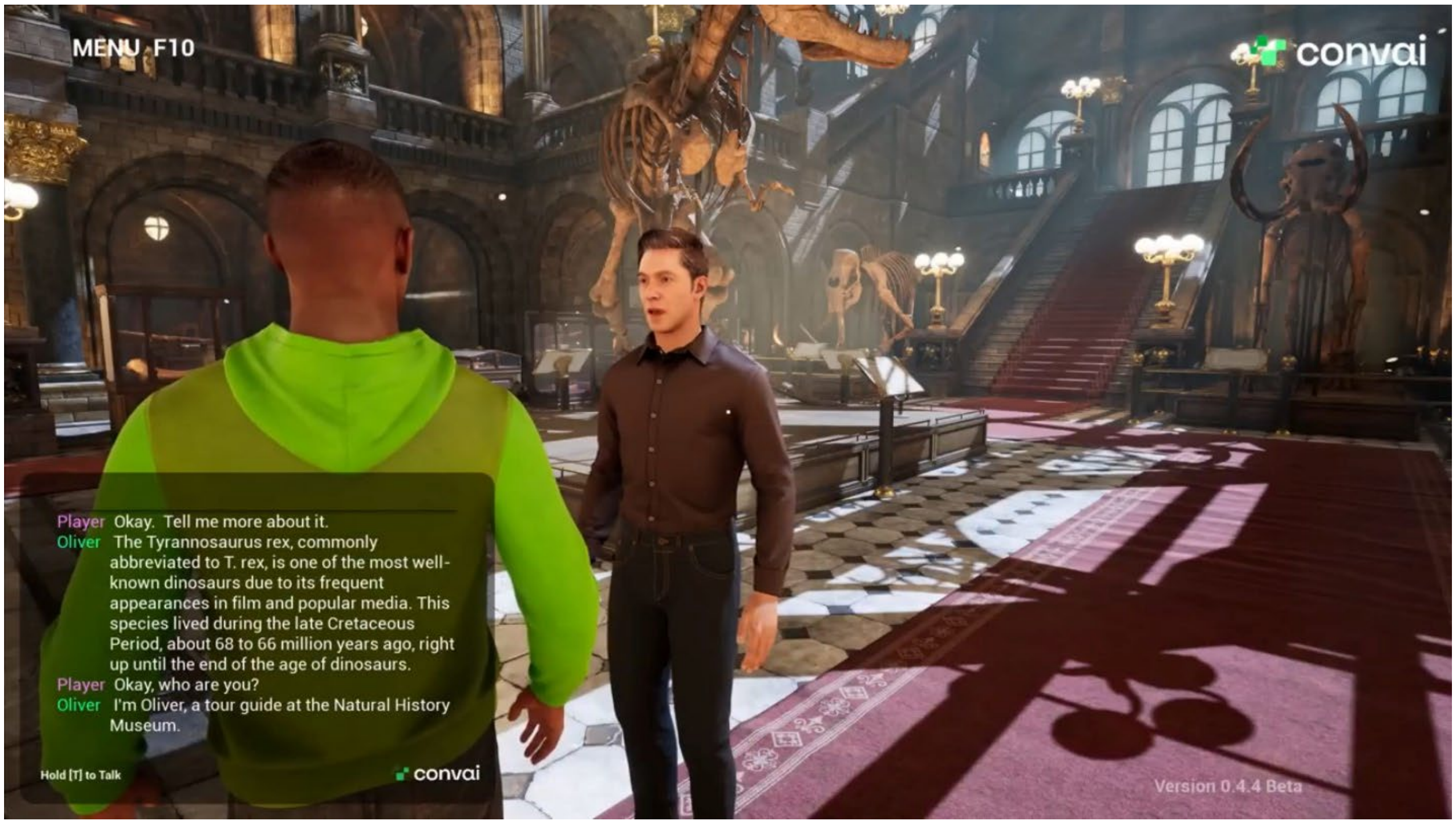

Figure 2. Illustration of an interaction of non-playable characters in a game using Interactive LLM Powered NPCs technology (Convai, 2023).

AI Town, developed by Convex.dev, is an innovative virtual town populated by AI characters who interact, chat, and socialize just like human characters (Convex, 2023). It combines artificial intelligence technology and creative programming to create a dynamic virtual community. In this simulation, every resident is an artificial intelligence entity with specific traits and behaviors, capable of engaging in conversations and social interactions. These AI residents inhabit a digital landscape that mimics real-life towns, complete with architectural and environmental elements (Figure 3). Users can visit AI Town and interact with its inhabitants in real-time by joining the conversation and immersing themselves in this digital environment. This interface provides a platform to observe AI behavior and language abilities in a social setting and explore advances in chatbot technology. AI Town offers an example into how research and testing of artificial intelligence could be done in a safe, controlled environment.

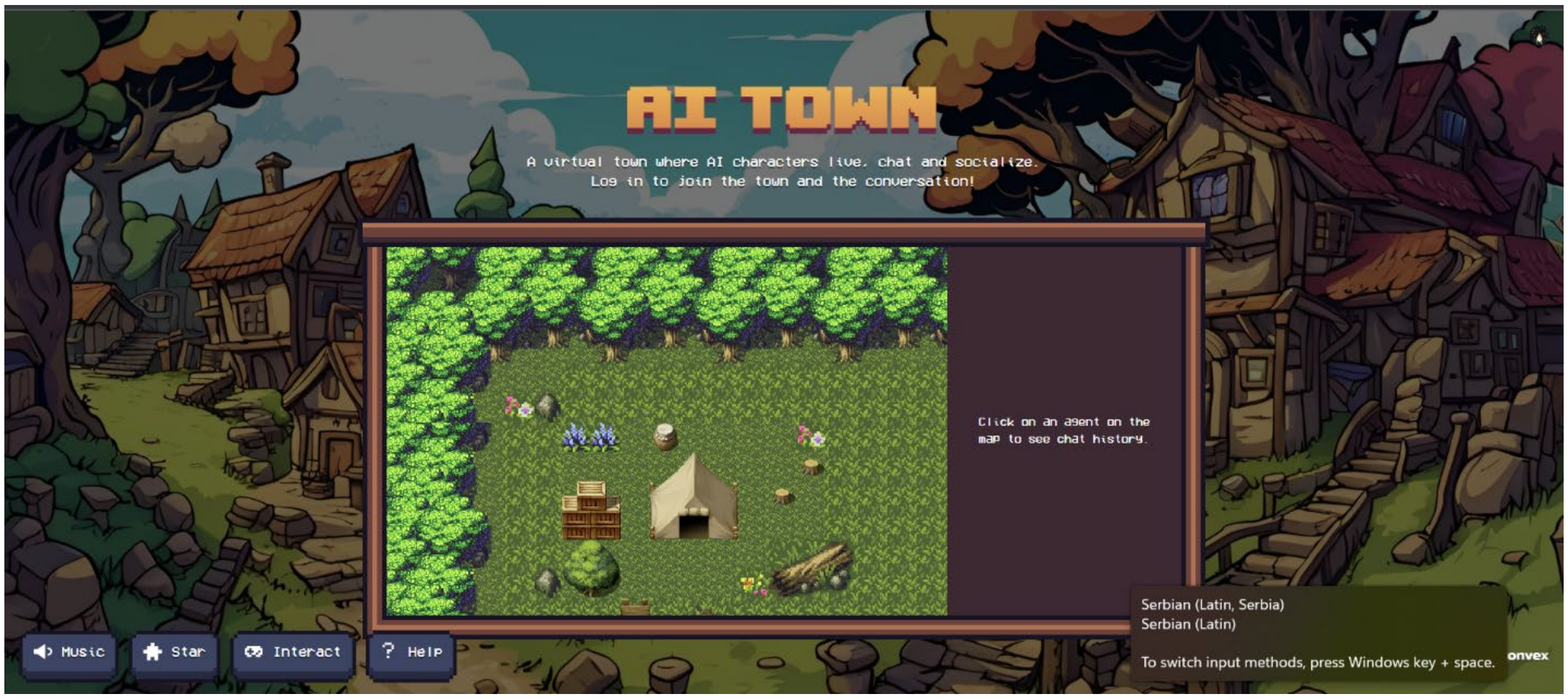

Figure 3. Screenshot of an AI Town web application (Convex, 2023).

*AI alignment*

The insurgence of AI options, while offering significant opportunities for both private and public sectors, have also presented a complex package of associated risks and ethical challenges. When considering previously developed AI systems, traditional methodologies of risk and error management have proven insufficient to mitigate potential harm efficiently and adequately (Brundage et al., 2018; Russell & Norvig, 1995). As a result, the growing body of research on AI safety necessitates an exploration of a multidimensional testing framework for these transformative technologies.

Prominent scholars have emphasized the risks associated with the unpredictability of AI systems. For instance, Irving & Askell (2019) pointed out that large-scale language models such as GPT can produce behaviors that, although desirable under controlled conditions, may present contentious outputs in unforeseen scenarios. On another note, Armstrong et al. (2012) expressed concerns on the “value alignment problem” with AI, stressing the necessity to hardwire ethical boundaries and human values into AI systems to avoid potentially catastrophic effects of an alignment failure.

Further exploring the domain of AI ethics, Véliz (2020) and Whittlestone et al. (2019) declared the importance of transparency, accountability, and public involvement in AI systems’ design and creation processes. Their work accentuates the need for democratized technology and calls for multiple stakeholders' contribution from different industries, sectors, and social-demographic backgrounds.

The incorporation of social science theories has gained ground in recent years due to their potential application in AI development. Rafols (2014) and Cave et al. (2019) suggested that taking lessons from social research methodologies could be a key ingredient to instill essential social sensibilities into AI systems, thus enriching their potential for more humane design. This marriage of technology and social sciences presents a promising avenue towards a socially-responsible AI future.

*Previous research*

The use of AI and specifically social simulation models in advancing scientific knowledge is an emerging area of interest in the field of AI. Social simulations as refuted machines have shown a significant contribution towards refuting prevailing theories in science, thus promoting scientific advancements (Mauhe et al., 2023).

The capacity of arguments in driving issue polarization has also been studied, indicating a potential role of AI in shaping public discourse (Kopecky, 2022). AI's role in polarizing debates among artificial agents can have implications for understanding political polarization in society.

The use of LLMs is also a promising direction in autonomous agent research. Wang et al., (2023) have shown that LLMs can acquire vast amounts of web knowledge and demonstrate human-level intelligence, indicating a potential role in diverse areas of social science, natural science, and engineering.

Simultaneously, the use of LLMs as a potential substitute for human participants in psychological research has also been evaluated (Dillion et al., 2023). Despite concerns about replicating human judgment, some studies show that LLMs can exhibit strong alignment with human judgments, thus suggesting that AI can play a role in replicating human subject studies in certain scenarios (Aher et al., 2023).

Furthermore, Generative AI is showing potential in improving social science research, online experiments, agent-based models, and content analyses (Bail, 2023). AI models can help in performing routine tasks, advancing programming skills, and writing more effective prose, which could transform the way social science research is conducted.

The potential of generative AI has been explored in strategic game experiments as well, with results suggesting that AI can generate realistic outcomes, and exhibit human-like behavior under appropriate conditions (Guo, 2023). AI's potential in mirroring human behavior has been studied through generative agents, interactive simulacra of human behavior (Park et al., 2023). Along these lines, AI has been utilized to create populated prototypes for social computing systems, showing potential in simulating real-world social interactions and behaviors (Park et al., 2022).

*Research objectives*

In light of latest developments in LLMs, the need for responsible AI, recent practical simulation-related applications and previous inquiries, this study aims to answer the following research question (RQ1): which theories and approaches derived from multiple disciplines could be useful as foundations for a coherent, multi-faceted simulation-based testing approach for AI?

This approach aims to create a more comprehensive and stringent process for assessing AI agents, particularly LLMs and within the specific context of a virtual reality framework that simulates life in a digital city. The increased stakes brought on by the sophistication of AI are seen as a pressing call to action for the design of more dependable and human-aligned models through this simulation-based approach (Dafoe, 2018; Bostrom, 2014), which could be useful for refining and fixing these systems (Mauhe et al., 2023).

We aim to delve into the depth of the opportunities and limitations provided by the merger of AI and various theoretical approaches, to nurture an effective, socially-responsible and comprehensive approach for testing and aligning LLMs within a digital environment (Russell & Norvig, 1995; Cave et al., 2018).

This paper tackles the novel and complex challenge of setting up a theoretical framework for testing and aligning LLMs. The subsequent sections will explore selected theories from various fields, their application to AI behavior and alignment, and their relevance in the context of a simulation-based approach. The latter part of the paper will delve into the practical aspects of applying these theories into a simulation-based approach to AI development. This includes the design and operation of a digital city populated by AI entities or 'digital citizens', their interactions and decision-making processes, and the resultant insights and applications. The paper concludes with reflections on the limitations of this approach and possibilities for future research. This exploration is vital in our relentless quest to ensure that AI technologies are effective, secure, and uphold the values of human society.

## Methods

In this research, two key methods were used to investigate the utility of a variety of theories from different scientific fields in the creation of a Digital City for AI testing and alignment. These methods include literature review and theoretical analysis.

The first method utilized in this research was an extensive literature review (Hart, 1998). A broad spectrum of scholarly literature was explored to obtain comprehensive insights into the field under study. These included academic articles, books, and reports covering Artificial Intelligence, LLMs, AGI, social robotics, social simulations, computer science, physics, and economics amongst others. The reviewed literature was selected based on its relevance, the prominence of the authors in the field, and the impact it has had in the scholarly community.

Literature on simulation-based approaches in AI testing and development was thoroughly explored to understand the current methods adopted by researchers and the challenges they face (Kelly et al., 2013; Bostrom, 2014; McEwan et al., 2015). The insights derived from the reviewed literature were instrumental in formulating the simulation design used in this research.

The goal of the literature review process was not only to understand the existing body of knowledge but also to identify gaps in the current research that our study could address.

The second method used in this study was theoretical analysis (Creswell, 2009). An extensive analysis was conducted to understand how theories from different fields could be applied to the development, testing and alignment of LLMs in a simulated digital city. This integrative and interdisciplinary perspective helped to develop the basis for the 'digital city' concept, highlighting the role these theories can play in AI alignment.

### *Procedures*

The process of identifying, assessing and narrowing down relevant disciplines and theories for inclusion in this research was a meticulous task. From the outset, the undertaking was centered on identifying those theories and constructs with the greatest relevance and application to the field of AI development and testing. The filtering process involved the application of a selection of critical criteria.

Firstly, the theories were assessed for compatibility with the overarching AI concept. A strong correlation with the principles of AI and the ability to enhance its understanding was a prerequisite. The theories also were evaluated based on their potential to contribute a unique perspective on AI behavior or its alignment, considering also cross-disciplinary connections.

The selection process was driven by a comprehensive review of literature in fields like computer science, psychology, sociology, and economics.

The critical aspect of this selection process was pairing the theories with appropriate methodologies or techniques. For instance, Matching the Big Five personality theory with computational personality recognition techniques for the creation of digital citizens or the application of game theory in concert with reinforcement learning approaches for AI decision-making.

The hierarchical clustering process based on theoretical importance, relevance, and application in the AI context was also utilized for selecting the theories and approaches. Hierarchical cluster analysis is a statistical method that builds a hierarchy of clusters iteratively. Starting with individual theories in separate clusters, in each iteration, we merged the two most similar ones until only one cluster was left (Everitt et al., 2011). Similarity was determined based on theoretical underpinnings, field of study, potential application in AI context, and overall compatibility with AI development philosophies. The dendrogram obtained from this analysis gave us an understanding of the interconnection and relatedness between theories, thus guiding our selection process.

Potential ethical and practical considerations relating to the application of theory to AI development were integrated into the discretionary process. Theories that highlighted ethical dimensions in AI development were seen as providing a cornerstone for the digital city simulation.

The efficacy of matching the theories with real-world machine learning techniques and their compatibility with the simulation design was considered. It was crucial to include the theories with relatively established techniques ensuring pragmatic application in the AI alignment process.

The process concluded with a peer review check, where subject matter experts reviewed our selection process and results to minimize biases and strengthen our chosen theoretical framework. This process of thoroughly reviewing each premise fed into our systematic selection process of evaluating the relevance and applicability of the theories in the context of AI.

The selection was then consolidated and constructed into a comprehensive framework for the digital city simulation. The entire process, from reviewing literature to choosing relevant theories, was iterative in nature, following the cyclic research model presented by Taylor et al. (2015). It involved constant reevaluation, feedback, and modification, ensuring that the theoretical framework was sound, relevant, and comprehensive.

From over 200 theories and methodologies identified in the early stages of the literature review, we whittled the list down to a focused beam of 10 theories and methodologies from various disciplines, which were deemed most relevant and promising for AI testing and alignment using a digital city simulation. This narrowed focus enabled us to make a deep dive into these select disciplines and theories, ensuring a detailed understanding and relevant application of each in the overarching design and operation of the digital city simulation. Over time, this iterative and detailed approach enabled us to create a unique, first-of-its-kind theoretical framework that seamlessly combines multidisciplinary findings to foster AI development, testing and alignment.

This multi-layered selection process utilized across 120 referenced studies and papers furnished the research with a well-rounded theoretical foundation. It also elevated the richness of the digital environment within our simulation approach while ensuring reliability within the context of AI testing and alignment.

**Establishing a framework for interactions in an autonomous digital city**

This section presents applicability of theories derived from sociology, biology, physics, social psychology and computer science to be used for the process of testing and aligning LLMs and AI. Complexities of LLMs and AI necessitate a rigorous, theory-based approach to aid in testing and alignment processes and to minimize potential risks (Rafols, 2014). Social science theories are valuable in understanding behavioral patterns and decision-making processes, potentially offering explanatory and predictive abilities for AI behavior (Ostrom, 2014). Subsequent subsections will offer an in-depth exploration of selected theories and their relevance within AI's context.

*The social simulation and reasoned action theories*

Integrating social theories in AI research provides perspectives for understanding AI behavior and alignment. Social theories focus on social relations, structures, and institutions that constitute society. They offer a theoretical lens for understanding social phenomena, behavior, and the intricacies of human interaction, which can be extrapolated for enhancing AI alignment. Particularly, the Social Simulation Theory and the Theory of Reasoned Action could be pivotal in this context.

The Social Simulation Theory stems from the broader spectrum of Computational Social Science (Cioffi-Revilla, 2014), emphasizing the power of computational methods to simulate, analyze, and draw insights from complex social phenomena. In the context of AI, this theory would be a key approach for testing LLMs. By simulating complex social dynamics computationally, researchers can create a more dynamic and realistic test environment, shedding light on how AI models might interact in various societal scenarios, and thus how to better align AI behavior to human values and expectations (Macy & Willer, 2002). This theory can contribute to the understanding of AI systems' behavior from multi-layered perspectives, which is crucial given the complex and often unforeseen consequences that AI systems can have in society (Helbing, 2015). However, the Social Simulation Theory also faces numerous challenges. One significant hurdle is the inherent unpredictability of real-world social systems. Simulating complex social dynamics in abstracted computational models inevitably involves simplifications, which can limit the accuracy and applicability of the results (Edmonds & Moss, 2005).

Originating from social psychology, Theory of Reasoned Action states that intentions drive individual behavior, formed by attitudes towards the behavior, subjective norms, and perceived behavioral control (Fishbein & Ajzen, 2010). While originally designed to understand and predict human behavior, the Theory of Reasoned Action may also be extended to autonomous agents in AI. It can guide the modeling of AI behavior in a virtual environment, thus influencing AI's intentions through programming norms and attitudes. By predicting and understanding the possible actions of AI, this theory could assist in aligning AI's actions with the regulatory norms and societal values (Sheeran & Webb, 2016). A major challenge posed by this theory is derived from the complex nature of emotions and irrational behavior, which greatly influence human decision-making but might be challenging to replicate in an AI. This complexity highlights the importance of multi-faceted approach to AI alignment.

*The Situated Action Theory*

The Situated Action Theory is an integral aspect of cognitive science that proposes a shift in viewpoint from the classic prescriptive comprehension of behavior to a more adaptive and situation-dependent one (Suchman, 1987). Applying this theory to AI development offers enhanced capabilities for AI behavior within their digital environments, thus making them more in sync with the dynamism and unpredictability of the real world.

Situated Action Theory contends that behavior is not just an outcome of premade plans but is spontaneously formed through continuous and dynamic interaction with the environment. In light of AI, this translates into AI models capable of proactive response modification as a result of changes in their environment rather than being solely driven by an extinctive set of actions (Varela et al., 1992). By doing so, we facilitate advanced AI systems that can independently make decisions based on the situation at hand within the parameters of a digital city, thereby being better equipped to navigate the inherent unpredictability present in large-scale neural networks (Cruz et al., 2023).

A situated cognition model offers a more dynamic outline for AI behavior by coupling the capacity to process information with appropriate autonomy to act in context. It emphasizes the need for cognition to be embedded in an understanding of, and interaction with, the environment and not just merely contemplative or inert (Kirsh, 2009). Drawing from cognitive sciences, computational approaches to situated cognition help analyze the interaction of AI with the environment, its dynamics, and adaptability to the perceived settings. Acknowledgment of perceptual-action loops can thus be considered as critical in implementing the Situated Action Theory in AI (Engel et al., 2013).

Translating these concepts into practical AI programming is a demanding task (Kirsh, 2009). Real-world factors are multifaceted and ambiguous, which may be difficult to replicate entirely within a digital environment. The dynamism involved in such a set-up would require models to be accurate enough to induce learning while being resource-efficient (Troitzsch et al., 1996). This dichotomy presents a major trade-off challenge needing careful consideration and smart solutions.

Though a formidable task, designing AI's adaptive behavior based on Situated Action Theory provides an avenue to unravel cognitive mechanisms in simulated environments (Hutchins, 1995). This paves the way for a sophisticated AI model that is not only capable of extracting information from its environment but can also adjust its behavior based on complex contextual information, strengthening alignment with human values, and contributing to secure, reliable AI systems.

Following the exploration of theoretical perspectives in Part I, the paper will expound on how we can apply the theories and insights from social science, robotics, and artificial intelligence into practical testing and alignment of AI. We will look at the creation of a 'digital city' and 'digital citizens' as a central aspect of our innovative simulation-based approach. Their interactions and decision-making processes within this simulated framework will form a crucial part of our study into autonomous behavior. Throughout, we will also discuss the valuable insights and practical applications of this approach, contributing to the refinement and alignment of AI models.

### *Complex Systems Theory*

In the quest to design realistic and effective simulations, the utility of Complex Systems Theory cannot be overlooked. As a computational and theoretical method, this approach helps to understand the behavior of systems characterized by intricate webs of interdependencies (Miller & Page, 2007).

Complex Systems Theory is founded on the primal assumption that emergent system behaviors can arise from simple local-level rules and interactions (Bar-Yam, 2003). Moreover, this theory particularly focuses on how small changes in the system can potentially herald large-scale effects - a characteristic often referred to as "sensitivity to initial conditions" or, more colloquially, the "butterfly effect".

The population of a city, be it real or simulated, shares numerous characteristics with complex systems. Both environments encapsulate localized systems or agents that independently follow simple rules but collectively generate emergent behavior at the city level. The inherent structures among these agents form a complex network, very similar to a real city that comprises various social, political, and economic networks interacting simultaneously (Batty & Torrens, 2001).

In the context of creating digital citizens in a simulated city, Complex Systems Theory is a pertinent guide. It can be harnessed in the development phase to design rules and behaviors for digital citizens. By underscoring the relations and dependencies between the different components or inhabitants of the simulated city, this theory can generate unique richness and complexities (Bar-Yam, 1997). Incorporating it benefits our understanding of societal phenomena, forms the foundation for testing the robustness of AI, and assists in aligning LLMs' behavior in digital citizens (Axelrod, 1997).

However, applications of this theory into AI development are not devoid of challenges. The notorious difficulty in predicting the complex systemic behavior and the associated implications demand a careful approach that combines continual monitoring, learning, and adjusting of the developed AI systems (Holland, 2006). This complexity accelerates the need to harness various theories, from social and robotics to psychology and game theory, creating a holistic approach to AI development.

### *Swarm Intelligence*

In running complex simulations involving hordes of autonomous agents, such as digital citizens of a simulated city, incorporating Swarm Intelligence can present distinct advantages. Swarm Intelligence, a subset of Artificial Intelligence, refers to the collective behavior of decentralized and self-organized swarming entities (Bonabeau, Dorigo & Theraulaz, 1999). This behavior, inspired by natural phenomena like fish schooling, bird flocking, and ant colonies, is characterized by collective intelligence that emerges from the interactions between simpler, individual agents (Kennedy, Eberhart & Shi, 2001).

In the context of creating digital citizens, Swarm Intelligence can assist in modelling and facilitating intelligent behaviour from a multitude of interacting entities. Agents can share local information and adjust their behaviors based on this shared knowledge, resulting in emergent global strategies that optimize simulated tasks (Karaboga & Akay, 2009). Examples abound in algorithms inspired by the behaviors of ants (Ant Colony Optimization algorithms), birds (Particle Swarm Optimization algorithms), and bees (Artificial Bee Colony Algorithms) that can be tailored to run complex simulations in a distributed, self-organized system (Blum & Li, 2008).

Furthermore, Swarm Intelligence offers crucial agents' patrolling abilities (Chevaleyre, 2004), a potential necessity in our digital city. For instance, agents can be designed to monitor and maintain the city's different sectors' security, following algorithms based on Swarm Intelligence dynamics.

But careful caution must be exercised in incorporating Swarm Intelligence, as transferring biological concepts of swarming behaviour to AI agents may not always manifest desirable results. Unforeseen system behaviours can emerge from Swarm Intelligence algorithms, yielding unexpected and undesirable outcomes (Kennedy, Eberhart & Shi, 2001). Further, the parameter selection for Swarm Intelligence algorithms can be a complex task due to the interconnected nature of the parameters (Clerc & Kennedy, 2002).

In spite of these challenges, Swarm Intelligence provides additional layers of complex behavior for the digital citizens and bolsters the interdisciplinary approach amalgamating social science theories, robotics, game, visual and complex systems theories. This comprehensive perspective creates a robust foundation for AI development, alignment, and research.

*The Multi-Agent System Theory*

The Multi-Agent System Theory embodies an important cornerstone in the conception and development of autonomous systems (Wooldridge, 2009). As we venture further into the domain of AI technologies, especially within the context of testing LLMs, understanding how multiple AI agents can work in conjunction or competition becomes increasingly critical.

Multi-agent systems are collections of several autonomous agents that interact with each other within a specific environment. These agents can be both cooperative and competitive (Russell & Norvig, 1995). This scenario is highly congruent with the simulated digital city environment envisaged for testing LLMs. In such an environment, each AI agent can be explored as an individual entity, having its unique attributes, characteristics, and decision-making abilities, according to the goals it has been programmed to achieve.

The theory provides us with insights into the potential interaction landscape of AI systems. By creating a multi-agent system, we allow AI models to interact among themselves as well as with the simulated environment, which can expedite the uncovering of emergent behaviors and systemic weaknesses or strengths (Mataric, 1998). With multiple AI agents, we can generate a range of multi-layered, dynamic scenarios that test the robustness and the adaptiveness of the AI models. This could lead to interesting observations about how AI agents learn to cooperate, compete, and negotiate, mimicking the dynamics of real-world complex systems (Shoham, 1993). These insights can lend crucial guidance in aligning AI decision-making processes to desired outcomes.

However, multi-agent systems are also not without challenges. The issues lying in the implementation of this theory could range from achieving synchronization among agents to dealing with conflicts and competition while reaching shared goals (Weiss, 2000). Furthermore, the architecture of AI systems in a multi-agent setup could swiftly evolve from complex to chaotic with the escalating number of agents. This complexity could make troubleshooting a significant challenge (Franklin & Graesser, 1997). There are also considerable technical hurdles in ensuring smooth, effective communication between agents in a multi-agent system (Lesser, 1999).

Despite these challenges, Multi-Agent System Theory holds substantial potential in shaping the future of simulation-based AI testing (Ossowski, 2013). Deeper understanding of this theory's implementation can enhance the precision and efficiency of AI models by providing a broader panorama of their potential interactions and making us better equipped to optimize their functionality.

## Creating elements of an autonomous digital city

Automated simulations, when performed in a digital domain, offer the potential for high reproducibility and scalability, mimicking complex, interactive scenarios within a digital city structure. They enable a systematic analysis of "real-life" scenarios. Massively complex systems can be simplified into essential actions and interactions for robust analysis and evaluation (Banks et al., 2000).

Our digital city employs a multi-agent-based simulation framework, which enables the modeling of a population of digital citizens (Bertacchini et al., 2013). Each digital citizen is an autonomous AI agent, modeled with a focus on social characteristics to increase the realism and complexity of interactions. These agent-based models offer an effective method for understanding complex environments - built on vast interactions among individual agents and their interactions with the environment. Indeed, the use of agent-based models has become commonplace across disciplines, from ecology to economics (Heath et al., 2019).

Creating a complex, interactive environment within a digital city requires a systematic and iterative process. Initial phases require the creation of autonomous agents, the digital citizens, that can operate within a defined parameter space (Silver et al., 2016). Silver et al. highlight that the success of these agents in the digital city directly depends on their autonomy level and the parameter spaces within which they operate. In our context, the agents are modeled on LLMs, giving them the capability to engage in sophisticated interaction, including natural language conversation.

The capacity to act on their own, in other words, their autonomy, defines the digital citizens' dynamics within the city (Lehman et al, 2018). Lehman et al. point out how autonomy has become a focal point in computing and robotics. It is crucial for the digital citizens to make decisions, conduct actions, and participate in the simulations dynamically.

One crucial aspect of developing the digital city is the environment's meticulous design, where the AI agents operate (Olson et al., 2013). The city should not be a mere backdrop; instead, it should function as a grounding influence, shaping the decisions and interactions of our digital citizens. Creating such an environment necessitates a detailed focus on various aspects, such as defining the interaction rules, constraints, and the potential choices for the AI agents (Olson et al., 2013).

In order to ensure continuity in the learned behavior and refinement of AI agents, Leike et al. (2017) propose a reinforcement learning approach. Reinforcement learning embedded in simulation allows for the automation of performance improvement, rewarding actions that lead to desired outcomes and penalizing those that do not. Given the high complexity of interactions and decisions occurring in our digital city, such reinforcement learning becomes crucial for developing AI's secure, value-aligned behavior.

The creation and successful implementation of automated simulations involve numerous critical aspects such as those explored in the subsequent sections: infrastructure, citizens, perception and cognition.

*Infrastructure through simulation engines*

A simulation-based approach to testing and aligning LLMs represents a critical leap in the quest of enhancing the reliability, performance, and security of AI. Kelly et al. (2013) propose an argument on the untapped potential of simulation-based approaches for AI-driven technologies. The authors assert that simulations offer a controlled environment where system behaviors can be analyzed under numerous scenarios, a concept that is crucial in the current study.

The proposed digital city mimics a 'real-world' environment but with controlled variables to assess the performance of AI agents and LLMs (McEwan et al., 2015). McEwan et al. emphasize the effectiveness of such environments in mimicking complex, interactive scenarios. They invite us to draw on the growing evidence that these simulation-based approaches are indispensable in testing AI systems.

This research uses a simulation-based scenario in a virtual reality framework, a paradigm that holds a growing influence in simulation studies (Barrett et al., 2022). Deploying a virtual reality framework rather than a basic 2D simulation offers an immersive and interactive platform. Coupled with artificial intelligence, the framework has an enormous potential for transforming the testing, design, and alignment of AI constructs (Vora et al., 2002). It offers an opportunity to critically investigate how AI agents interact in life-like scenarios, therefore providing a tool for understanding autonomous behavior.

The incorporation of a virtual reality framework into the simulation approach encompasses a broader context for the use and amplification of AI. As Rouse et al. (1992) highlight, advancements in virtual reality have a significant influence on artificial intelligence. The use of virtual reality facilitates the concept of 'digital twin' or a 'mirror world' – an almost realistic, virtual representation of the real world that allows for dynamic interaction and learning (Bolton et al., 2020; Bojic, 2022). The creation of a virtual environment that simulates a realistic city is not only beneficial for the research on AI behavior, but also opens up possibilities for future anthropological, sociological, and psychological studies (Sun, 2005). A virtual reality-based city could also be used for understanding social behavioral dynamics. As a result, the efficacy of this method in yielding comprehensive and cross-domain insights is undeniable.

A simulation-based approach is a valuable tool for observing, evaluating, and aligning the behavior of AI agents, particularly LLMs through a more realistic and interactive approach (Bostrom & Yudkowsky, 2014). This approach, underpinned by a virtual reality framework for simulations, could potentially revolutionize our understanding and approach towards advancing AI.

Developing an immersive simulated environment, robust in its complexity and capable of hosting myriad digital citizens, necessitates the integration of powerful simulation engines. These engines, a collection of complex algorithms and computational models, serve as the backbone of the simulation, allowing for the creation, interaction, and evolution of entities within a virtual environment (Smith, 2014).

Simulation engines offer immense flexibility and control, allowing researchers to design varied scenarios, inject contingencies, and monitor the system's evolution in real-time. Several game-based technologies such as the Unity Engine (Unity Technologies, 2020), Unreal Engine (Epic Games, 2020), and Godot Engine (Godot Engine contributors, 2020) possess powerful physics, graphics, and AI capabilities that render them fit for creating interactive, immersive environments with realistic physics.

Unity, for example, allows researchers to create diverse, visually rich environments and control them at both macro and micro levels. It supports scripting in C#, facilitating the implementation of unique AI logic to control digital citizens within these environments. The Unreal Engine, on the other hand, is known for its sophisticated visual rendering capabilities, ideal for creating realistic, high-detail environments. It provides native support for Behavior Trees, a tool used to design complex AI behavior (Rabin, 2014).

While game-based engines are more suitable for visual simulations, other specialized simulation environments such as MATSim (Multi-Agent Transport Simulation) and MASON multiagent simulation toolkit are better suited for large-scale urban simulations and social dynamics.

Adoption of these engines comes with its challenges, including the need for specialized knowledge and potential hardware limitations. However, the benefits, such as the ability to observe an AI's behavior in a controlled environment, far outweigh the learning curve and resource demands (Yan et al., 2023).

Combined with robust theories from social sciences, robotics, game theory, visual and complex systems, and swarm intelligence, simulation engines serve as pivotal tools in developing robust, versatile, and impactful AI.

### *Citizens - The Big Five theory*

Shaping the digital citizens serves a central role in our simulation of a digital city. It is these digital citizens, the primary agents, exhibiting high degrees of autonomy that bring life to the city and the scenarios unfolding within it. In essence, the behaviors, interactions, decisions, and profiles of these AI agents determine the richness and complexity of the simulated environment (Bartneck et al., 2015).

Digital citizens can be viewed as AI actors, playing the part of the city's inhabitants, depicting complex behaviors that could resonate with the population of a real city. By endowing the agents with distinctive characters, norms, and behaviors, we mimic a community's diversity in a real city, thereby enriching the simulation's scenarios and insights.

Creating these digital citizens demands an intense focus on the autonomy level. Autonomy, defined as the capacity to act on one's own, forms the crux of these AI agents' dynamics (Stone & Veloso, 2000). Stone and Veloso emphasize that portraying autonomy in robotics or AI systems is critical to determining the value and effectiveness of their performance. High-level autonomy of these LLMs represents both an algorithmic challenge and a conceptual step forward, allowing reactions and proactive engagement with unforeseen events in the digital city to be studied (Stone & Veloso, 2000).

Understanding and effectively replicating the depth and complexity of human personality and behavior is a non-negligible aspect of creating digital citizens within our simulated environment. To this end, delving into the annals of psychology theories can provide the necessary scaffolding. These theories, such as The Big Five personality traits theory, are remarkably useful in devising the specifics of distinct, unique personalities and behaviors within our digital citizens.

Originally conceived by Goldberg (1990) and elaborated upon by Costa and McCrae (1992), The Big Five theory suggests five broad dimensions of human personality: Openness, Conscientiousness, Extraversion, Agreeableness, and Neuroticism. Each dimension represents a spectrum, with individuals falling somewhere along this spectrum, thereby shaping unique personality profiles. This model has gained significant acceptance in psychology due to its comprehensiveness and empirical support (John, Naumann, & Soto, 2008).

In the context of creating digital citizens, the Big Five theory could potentially govern the nature and degree of variance among the inhabitants' personalities in our simulated city. Digital citizens' behaviors can be programmed based on combinations of these five dimensions, thus achieving diversity in behavioral patterns akin to a real-world city populace. This not only enhances the realism and richness of the simulation but also provides a structured means of attributing coherent, consistent behavior patterns to individual AI agents (Yarkoni, 2010).

Applying such psychological theories to AI also involves the task of embodying a breadth of human emotions and motivations, such as intricacies of emotional intelligence and social cognition. Models like the Theory of Mind, which reflects the ability to attribute mental states to oneself and others (Premack & Woodruff, 1978), can enhance digital citizens' interactivity, leading to a more dynamic and authentic simulation. Humans use the Theory of Mind in everyday interactions and empathic understanding, and incorporating this element into AI systems could allow them to predict and respond more flexibly and naturally.

To accurately implement psychology theories in digital citizens, research from computational personality recognition could be utilized, wherein machine learning and natural language processing techniques are used to detect and assign appropriate personality characteristics (Van Pinxteren et al., 2020).

Several different learning models can be implemented in creating digital citizens. For instance, reinforcement learning models the dimension in which the AI evolves as each digital citizen learns from interactions with the environment and other agents (Sutton & Barto, 2018). It's a powerful, flexible paradigm for defining the learning of digital citizens, thereby encouraging consistently increasingly value-aligned actions (Sutton & Barto, 2018).

The actions of the LLMs in the virtual city comprise one aspect of our scenario simulation. Equally important is their ability to interact, to respond and engage with their environment, other AI agents, and external inputs (Zhang et al., 2018). Zhang et al. point out how advanced machine learning techniques can be employed to develop conversational capabilities, further improving the simulation's reality.

These AI agents, significantly enhance the potential of running multi-agent simulations, including collective decision-making scenarios, cooperation, competition, and conflict (Leibo et al., 2017). Consequently, this results in a higher understanding of the complex social and interactive scenarios that we might encounter in the real world.

Digital citizens' personification allows for a gravity of interaction and personalization that strengthens the depth and breadth of the simulated city. Equipped with a high degree of autonomy, natural language-processing capabilities, character traits, and unique behaviors, these AI agents form the base of our simulated environment and are necessary to fulfill the objectives delineated for AGI development (Bostrom, 2014).

*Perception – computer vision approach*

The perceptual abilities of digital citizens feature prominently in the realization of an immersive and engaging simulation environment. To this effect, the incorporation of computer vision is paramount. Computer vision, a critical field within Robotics and AI, involves the automatic extraction, analysis and understanding of useful information from images or video sequences (Szeliski, 2010). Consequently, computer vision can provide digital citizens the ability to perceive and interpret their environment.

Several theories and techniques within computer vision can be leveraged to allow AI agents to distinguish between objects and individuals, identify patterns, interact more naturally or even anticipate possible future states of the environment. One such technique is Deep Learning, a class of machine learning algorithms that has been extremely effective in the realm of Computer Vision (LeCun, Bengio, & Hinton, 2015). Convolutional Neural Networks, a variant of deep learning, can aid digital citizens in identifying and categorizing the wide range of visual stimuli typically present in a cityscape (Krizhevsky, Sutskever, & Hinton, 2012).

Semantic segmentation, an advanced computer vision technique, may be used to enable AI agents to understand the varying components of their surroundings better, by classifying different parts of images into distinct categories (Long, Shelhamer, & Darrell, 2015). This can make for more sophisticated navigation and interaction within the digital city, setting the stage for more realistic scenarios.

Implementing concepts from Visual Scene Understanding, an area of research studying how observer characteristics, such as familiarity, guide anticipation and action planning (Henderson and Hayes, 2017), could also grant our AI agents the ability to make complex decisions based on the visual information they process.

While integrating computer vision into digital citizens, it is essential to bear in mind the potential pitfalls, such as the generalizability of training to different settings (Torralba & Efros, 2011) and the problem of 'adversarial examples' where small input modifications can make AI outputs incorrect (Szegedy et al., 2013). But despite these challenges, the blend of AI's perceivable ability with behavioral autonomy afforded by psychology, game and social theories, could significantly enhance the richness, realism, and utility of the simulation, pushing the boundaries of AI development.

*Cognition – game theory*

The richness of the digital environment within our simulation approach allows one to manipulate numerous variables, thereby observing the corresponding influence over the AI agent's behavior. The interactions and decision-making capacity of our digital citizens arise as focal points of this study (Turing, 1950). Turing suggests that we can attribute intelligence to an entity by observing its ability to make "reasoned" decisions under changing circumstances and its capacity to interact convincingly.

One of the unique traits exhibited by our digital citizens is their capacity to engage in interactions. These interactions could arise between multiple AI agents (multi-agent interaction) or between AI agents and human users (Wilks, 2010). Wilks explains that interactions with multiple agents can lead us to comprehend how complex social dynamics play out. In the context of our AI-driven virtual city, these interactions provide the primary source of data, providing insights into the behavior of AI agents in various scenarios (Russell et al., 2015).

Interactions in a virtual environment can vary in numerous ways, from simple exchanges – such as greetings – to more complicated ones involving conflict resolution and cooperative tasks (Jennings et al., 1998). Jennings et al. emphasize the need for a well-defined protocol guiding such interactions, critical in ensuring the scalability and success of multi-agent systems.

As for decision-making, they are crucial in an autonomous agent's capacity to take independent actions. These decisions could range from simple binary choices to complex resolutions involving trade-offs between conflicting interests or values (Sutton & Barto, 2018). Sutton and Barto highlight the role of reinforcement learning in shaping an agent's decision-making process - allowing it to 'learn' from its past actions and outcomes.

The application of game theory stands out as a significant tool that provides valuable perspectives for understanding and ultimately influencing the behavior of LLMs. Game theory, originally derived from economic and mathematical contexts, is a well-established theoretical framework that describes and analyzes decision-making scenarios where the outcomes for each participant depend upon the actions of all (Fudenberg & Tirole, 1991). In the AI landscape, game theory's main draw is its power to model strategic interactions between multiple agents (Shoham and Leyton-Brown, 2008). In an AI ecosystem constituting multiple self-learning agents, each AI's decision is inherently interdependent, affected by the decisions of others in the same ecosystem. This presents an inherent multi-agent coordination problem. Game-theoretic strategies can provide solutions to this problem by providing a mathematical formalization and an analytic platform for these dynamics, thereby fostering cooperative behaviors (Busoniu, Babuska & De Schutter, 2008). For example, in networked AIs, game-theoretic tactics like the Nash Equilibrium can help align the AIs towards common objectives, by attaining a stable state wherein no AI can gain by unilaterally deviating from its chosen strategy (Chalkiadakis, Elkind, and Wooldridge, 2011).

Game theory contributes to the field of machine ethics, providing a structured methodology for AI to make ethical decisions when pitted against dilemmas. The incorporation of game theory in the design focuses on actual decision-making processes rather than imbuing AIs with abstract, universal principles. By assigning utilities to each possible outcome, ethical AIs can navigate through complex decision trails, evaluating and selecting the most ethical course of action (Etzioni, 1990).

The interactions and decisions of our digital citizens within this digital space create a rich source of data for multiple applications, from refining future iterations of these AI models to informing policy decisions in digital technology use (Amodei et al., 2016). Amodei et al. reiterate the significance of the decision-making models, especially in unpredicted scenarios, to enhance our AI agent's ability to align with human values.

It is essential to observe that these simulated interactions and decision-making scenarios remain subject to ongoing refinement and calibration, ensuring they accurately reflect potential real-world situations (Amodei et al., 2016). Amodei et al. highlight the need for adopting an iterative approach in refining and perfecting the simulations. This perspective resonates with the developing nature of AI.

Interactions and decision-making within our digital city offer a detailed, real-time understanding of how autonomous agents respond to a variety of scenarios. The evolution of strategies and choices reflects the trends and challenges that might be encountered, thereby advancing our ability to guide AI development towards safety and alignment.

## Discussion

### *Valuable insights and practical applications*

The application of an innovative simulation-based approach to study LLMs in a digital city provides critical insights that hold valuable implications for AI and its expansive usage. Such an approach can contribute to the development and refinement of AI by providing the opportunity to observe and correct anomalies or undesired behaviors in a controlled, reusable, and adaptable environment (Shoham et al., 2018).

According to the above literature, the project of an autonomous digital city for AI testing may consist of multiple modules, such as infrastructure (simulation engines) and citizens (personality, perception, and cognition). The infrastructure module provides a controlled environment where AI behaviors can be analyzed and tested in numerous scenarios. A simulation engine utilizes advanced algorithms and computational models to create, interact with, and evolve entities within a realistic virtual environment. The Citizens (AI Agents) module simulates city inhabitants, exhibiting complex behaviors and a high degree of autonomy. The Computer Vision and Perception module grants AI agents the ability to perceive and interpret their environment, improving realistic scenarios. The Cognition and Decision-making module emphasizes the autonomous decision-making capabilities of AI agents, allowing them to interact and adapt to their environment.

By deploying LLMs in a virtual reality environment, we emulate life via an imagined space equipped with digital citizens. Here, potential interactions and scenarios are virtually limitless, encompassing casual daily interactions, complex social situations, and even unexpected, novel circumstances (Castelfranchi, 2000). This immersive, controlled environment allows for simulations that hold significant anthropological, sociological, and psychological value in addition to their technological implications (Castelfranchi, 2000).

The generated knowledge extends beyond the informational dimension of AI performance and alignment. These insights can be applied to forecast, understand, and even shape the potential effects of AI implementation in different societal sectors (Mnih et al., 2015). The potential effects of AI integration into areas such as healthcare, education, governance, and commerce can be tested, adjusted, refined, and optimized within the digital city environment before actual implementation (Mnih et al., 2015).

The diverse data generated within the digital city provides valuable ground for addressing both overt and subtle biases that may become embedded within AI systems (Caliskan et al., 2017). The analysis of AI agent interactions can aid in identifying and correcting these biases to ensure AI deployments remain equitable, efficient, and reflective of the diverse values of humanity at large (Caliskan et al., 2017).

The specific domains where AI can bring transformative change – such as autonomous vehicles, robotics, customer service, and language translation – can also benefit from the data and knowledge generated from our approach. The observed patterns and anomalies in the digital city can inform us about potential obstacles, improvements, misalignments, and benefits that AI might face in these specific domains (Esteva et al., 2019).

The digital city as a research environment also serves as a crucible for AI's ethical and moral considerations. As the digital citizens interact, make decisions, and evolve, researchers can gain rich insights into the ethical boundaries and value alignment challenges associated with AI (Bostrom, 2014). The novel application of a simulation-based approach within a digital city offers a vital pathway for the sustainable and beneficial development of AI.

The key point would be the autonomy of different agents within the digital city. That means agents would be able to move through the city as they wish, engage in activities of their own choosing, and interact with other agents with whom they share common goals or interests. Conversation between agents would be just one possible activity. The goal would be to increase the complexity of the autonomous city, thus making it more realistic as it evolves into a playground for AIs. Having multiple AI agents participate in the digital city through their agents could be useful to establish testing and correction mechanisms. While more AIs would interact, trustworthy ones could serve as teachers and controllers of those that need to be tested and further aligned to human uses, needs, and values. An illustration of this could be seen in Figure 4. By fostering a deeper understanding of AI dynamics in a variety of scenarios, society can better seize the opportunities and manage the risks associated with AI innovations.

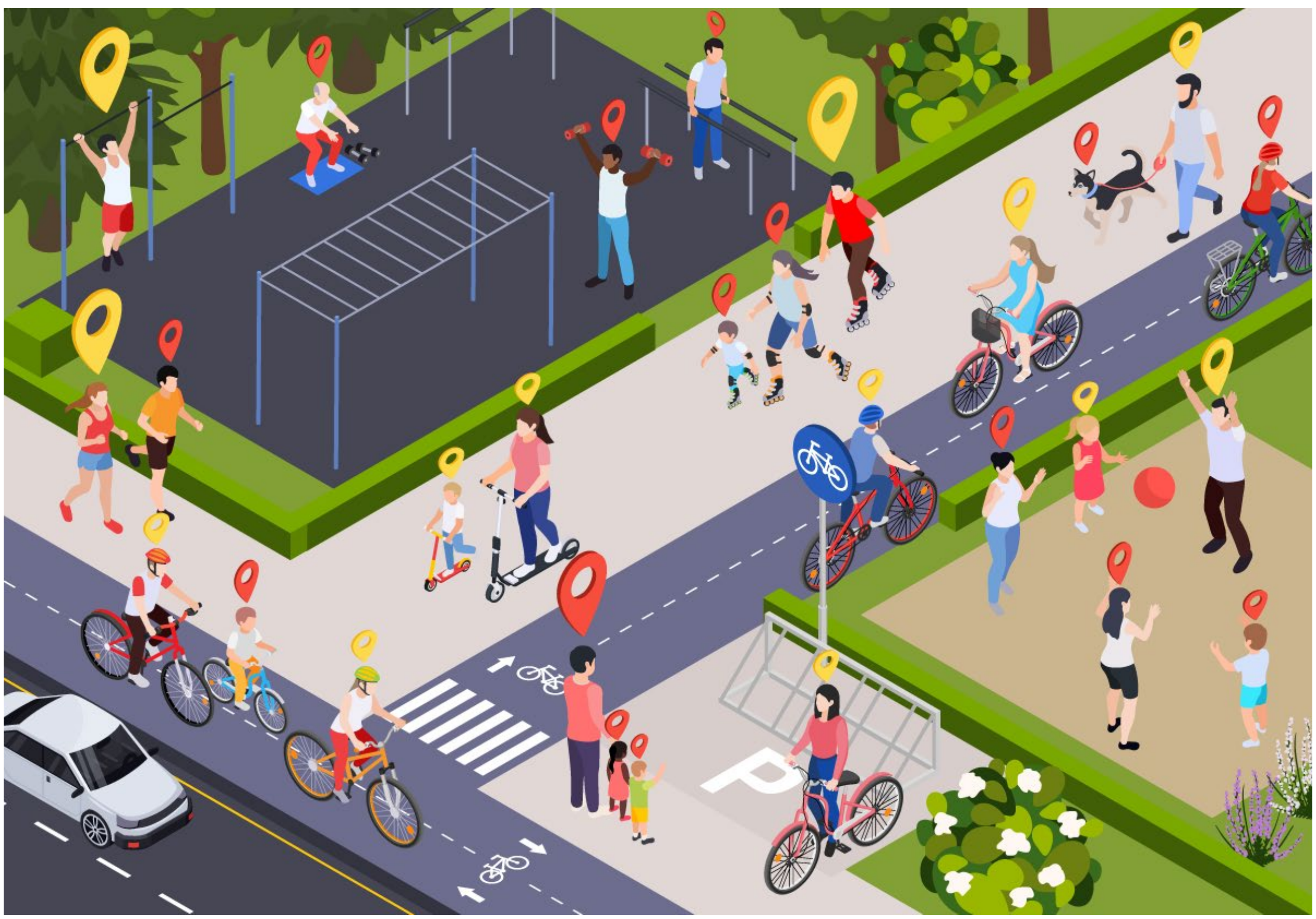

Figure 4. In the dynamic digital city, numerous autonomous agents, represented as distinctive yellow and red icons, serve as simulations of AI. This digital city is constructed with diverse structural elements such as buildings, parks, roadways, and transportation vehicles, reflecting the diverse facets of a real metropolis. The autonomous agents move around and interact within this landscape, demonstrating active engagement and reflecting the breadth of potential AI actions and interactions. These agent movements and interactions are indicative of continuous learning, decision-making, and evolution, which are inherent aspects of AI. This complex, multi-agent system within a digital city serves as a critical testing and alignment ground for AI development, capturing the numerous opportunities and complexities seen in AI testing and alignment.

*Refinement and alignment of AI Models*

Fine-tuning and aligning AI models to match human values is one of the most pressing needs of the high-tech world today (Bostrom, 2014). This is particularly important when reshaping AI, which could potentially replicate or even surpass human-like reasoning and cognition, including ethical and moral decision-making (Bostrom, 2014).

The innovative simulation-based approach facilitates the observation and correction of anomalies and misalignments of AI models. Observations from their interactions within the digital city can guide AI developers to identify areas of improvement, thereby achieving a higher degree of alignment with human values (Russell et al., 2015).

One important factor in AI alignment is understanding how the models generatively encode knowledge and concepts from the data they are trained on (Ring & Orseau, 2011). Simulation-based testing can offer insight into whether the AI model cognitively 'understands' what it has learnt, allowing us to understand its decision-making processes better (Ring & Orseau, 2011).

Decision making in AI is guided by reinforcement learning, which is naturally built into our simulation approach (Sutton & Barto, 2018). While reinforcement learning remains a powerful tool, it is necessary to carefully manage this learning process to avoid reinforcing undesired values or behaviors inadvertently (Amodei et al., 2016). As Amodei et al. observed, any approach intending to align AI with human values must consider the complexities of human value systems and the potential for unintended consequences.

In our simulation approach, the behavior of digital citizens within the digital city provides a rich data set for analysis. This data can be utilized to train AI models to act in ways that adhere to acceptable norms and societal values (Christano et al., 2017). The unique, scenario-based insights generated within the simulations can further inform the development of robust safety measures and the alignment of AI goals with broadly accepted human values (Christano et al., 2017).

The wealth of behavioral data and interaction analysis gathered through the simulation approach allows for unprecedented refinement of AI decision-making, leading to safer, controlled, and value-aligned AI systems. These insights can aid AI researchers and policymakers in mitigating the risks and maximizing the benefits of AI integration into society. Also, some of the theories discussed in this study could be used to address specific aspects of alignment, as indicated in Table 1.

Table 1: Ethical considerations and theories informing autonomous AI testing and alignment

| Ethical Considerations in Autonomous AI | Corresponding Theoretical Framework |
| --- | --- |
| Ethical Alignment | Social Simulation Theory |
| Controllability | Theory of Reasoned Action |
| Predictability | Situated Action Theory |
| Unpredictability in real-world dynamics | Complex Systems Theory |
| Autonomy in decision-making | Swarm Intelligence |
| Multi-layered AI behaviours | Multi-Agent System Theory |
| Autonomy and Interaction abilities in AI | Perception and Cognition in AI |
| Security and Safety in AI | Game Theory |

*Repercussions for the simulation argument*

The successful production of an artificial world containing independent, intelligent, and interacting digital entities naturally evokes contemplation on the simulation argument. This hypothesis, originally proposed by philosopher Nick Bostrom, suggests that advanced civilizations could possess the technology to produce realistic, convincing simulations of past eras peopled by conscious digital entities (Bostrom, 2003). The development of a sophisticated simulated environment could bear significant implications for such a hypothesis and pose intriguing philosophical questions.

The creation of a rich simulated city resided by autonomous digital citizens provides a tangible example of the feasibility of developing such a system, supporting the theoretical possibilities postulated by the simulation argument. Although our technology is primitive compared to the advanced civilizations in Bostrom's hypothesis, our progress underscores the potential reality that future technological advancements might enable the realization of full-scale, hyper-realistic simulations (Schneider, 2008).

In this context, the possibility that we might unknowingly be part of such a simulation ourselves becomes a more conceivable prospect. Our ability to create simulated environments that imbue digital citizens with some form of perceived consciousness could mirror an advanced civilization's ability to do the same on a magnitudinally larger scale. As such, there arises the existential question of whether our reality is, indeed, 'real' or merely a highly sophisticated simulation (Chalmers, 2010).

Validating or debunking the simulation hypothesis remains a vexing challenge. Current scientific methodologies fail to offer any verifiable means to do so. In fact, one argument suggests that if we are living in a perfect simulation, we may never be able to discern our reality's true nature (Bostrom, 2003).

Expanding our understanding and capability of creating simulated realities should be accompanied by ethical considerations, philosophical reflections, and a commitment to uncovering verifiable truths about our world, simulated or otherwise. Ongoing technological advancements reinforce the plausibility of the simulation hypothesis, impelling continuous exploration of our existence's nature and purpose in increasingly uncertain landscapes.

## Conclusion

The accelerated growth of Generative AI, and notably, LLMs, necessitates innovative approaches to ensure security and alignment of these models with human values. An effective way to achieve this is through the development and implementation of simulation-based methodologies. Through the application of various social sciences and robotics theories, considerations of computational social dynamics, human behaviors, ethics, and perception can be thoroughly examined, thereby providing crucial insights into AI behavior in the context of diverse societal scenarios.

To answer the research question posed in this inquiry (RQ1), the application of theories and approaches derived from multiple disciplines to create a comprehensive, multi-faceted simulation-based testing approach for AI encompasses several areas:

From Sociology, Social Simulation Theory and Situated Action Theory can aid in creating AI simulations reflecting human social behaviors and conditions under which actions occur. Theory of Reasoned Action derived from Social Psychology could be used to simulate human decision-making processes within AI, reflecting how individuals process information and make decisions. From Physics, Complex Systems theory can be used as a backdrop for creating AI testing simulations that bear the components of complexity, dynamism, and interconnectivity. Swarm Intelligence rooted in Computer Science and Biology, could be used to replicate collective behavior of decentralized, self-organized systems within AI. Multi-agent System Theory from Computer Science could be essential in building AI simulations where multiple agents interact with each other, mimicking real-world multi-actor scenarios. Utilizing Big 5 Personality theory from Psychology could assist in creating AI that simulates human personality characteristics, enhancing its level of realism and relatability. Computer Vision, a branch from Computer Science, could form the basis for visual processing and recognition capabilities of AI. Game Theory from Economics can construct simulations where AI tests strategic interactions between rational decision-makers. Lastly, Simulation Engines, another facet of Computer Science, are fundamental in creating the simulations upon which AI operates. All noted theories and approaches are listed in Table 2 with adequate references.

These multidisciplinary theories and approaches can collectively construct a robust, versatile simulation-based testing environment for AI, augmenting its adaptability and authenticity in various contexts.

Table 2: Theories and approaches identified as the most adequate for simulation-based AI testing and alignment framework

| **Aspect** | **Approach** | **Field** | **References** |
|---|---|---|---|
| **Interactions** | Social Simulation Theory | Sociology | Cioffi-Revilla, 2014; Macy & Willer, 2002; Edmonds & Moss, 2005 |
| | Situated Action Theory | Sociology | Suchman, 1987; Varela et al., 1992; Kirsh, 2009\| |
| | Theory of Reasoned Action | Social Psychology | Fishbein & Ajzen, 2010; Sheeran & Webb, 2016\| |
| | Complex Systems | Physics | Miller & Page, 2007; Bar-Yam, 2003; Bar-Yam, 1997; Axelrod, 1997 |
| | Swarm Intelligence | Computer science and biology | Bonabeau, Dorigo & Theraulaz, 1999; Kennedy, Eberhart & Shi, 2001; Karaboga & Akay, 2009 |
| | Multi-agent system theory | Computer science | Wooldridge, 2009; Russell & Norvig, 1995; Lesser, 1999 |
| **Agents** | Big 5 personality | Psychology | Goldberg, 1990; Costa & McCrae, 1992; Yarkoni, 2010% |
| **Perception** | Computer Vision | Computer science | Szeliski, 2010; LeCun, Bengio, & Hinton, 2015; Krizhevsky, Sutskever, & Hinton, 2012 |
| **Cognition** | Game Theory | Economy | Fudenberg & Tirole, 1991; Shoham and Leyton-Brown, 2008% |
| **Infrastructure** | Simulation Engines | Computer science | Unity Technologies, 2020; Epic Games, 2020; Godot Engine contributors, 2020 |

The creation of a controlled and reproducible environment, through the design of a digital city populated with digital citizens, allows researchers to observe and evaluate their behavior under various conditions. Social simulation and theory of reasoned action provide the foundation of our approach, framing the study of digital citizen behavior within the context of social relations.

In an effort to effectively manage the often complex and unpredictable behaviors of AI technologies, the application of various theories and approaches is crucial. The former aids our understanding of the interactions and dynamics between AI agents, while the latter allows AI to adapt its behavior in reaction to changes in its environment. Both theories contribute significantly to the alignment of AI decision-making with desired outcomes, despite the challenges that may arise in the process.

Our innovative approach also emphasizes the development of automated simulations, allowing the behavior of LLMs to be studied exhaustively within the digital city. The use of autonomous digital citizens and the exploration of their interactions and decision-making on a large scale provide a robust framework for understanding autonomous behavior. Not only do these insights inform the refinement of AI, but they also hold significant potential for broader applications across various sectors of society.

Equally critical in this process is the refinement and alignment of AI models to propagate secure, controlled, and value-aligned AI systems. The digital city simulation offers an environment where AI can evolve and be increasingly value-aligned with each iteration. However, the complexities inherent in translating the theories into practical AI programming and replicating real-world effects within an artificially confined environment present new challenges - challenges to be addressed with iterative refinement and disciplined learning and development.

This study demonstrates the potential of a simulation-based approach in testing and aligning AI with human values. Nevertheless, it also reveals the complexities and challenges intrinsic to this process, emphasizing the importance of ongoing refinements in theory, design, and application. As we delve into the era of AI and beyond, the steps we take must be exploratory in nature, continually acknowledging the dynamism of the terrain ahead. Through such constant engagement and innovation, we can aspire to design AI technologies that are not only effective and secure, but also respect and uphold the values of the human society in which they operate. The insights offered in this research illuminate a potential path – though perhaps initially challenging, it holds the promise of a secure and value-aligned future for AI.

*Similarities with CERN*

The theoretical framework presented in this paper to create a simulation of a digital city for AI testing and alignment shares similarities with CERN (2022), the European Organization for Nuclear Research, in terms of its large-scale, multidisciplinary nature. Like CERN, it's a significant scientific effort that requires collaboration from various fields. However, there are also notable differences. While CERN is focused on advancing our understanding of the fundamental laws of nature through experiments in high-energy particle physics, the digital city simulation aims to advance AI by providing a controlled environment where AI agents can interact, learn, and evolve. The purpose of this simulation is to provide insights into the behavior and decision-making processes of AI agents, thereby helping to refine and align AI technologies.

CERN relies on physical infrastructure and real-world experiments, whereas the digital city simulation is entirely virtual, relying on AI models, algorithms, and computational resources. Although both CERN and the digital city simulation involve complex, dynamic systems, the nature of these systems differs. CERN investigates the behavior of subatomic particles, while the digital city simulation investigates the behavior of AI agents within a complex, simulated societal environment. Both CERN and the digital city simulation represent grand scientific efforts aiming to improve our understanding of complex systems, whether they are natural or artificially designed. They utilize state-of-the-art technology and draw on a broad range of scientific disciplines, showcasing the power and value of interdisciplinary collaboration in scientific research. They both represent ambitious, forward-looking projects that have the potential to make significant contributions to their respective fields of study.

*Limitations of the study*

Despite the innovative approach adopted in this study, its limitations must be acknowledged. Firstly, the utility of a simulation-based approach relies heavily on the accuracy and complexity of the simulation or the modeled digital city itself. There is an inherent challenge in replicating the multifaceted features and unpredictability of the real world within a virtual framework. While efforts have been made to ensure a broad range of scenarios and interactions within this study's digital city, the certainty of covering all potential variables and eventualities remains elusive.

The adoption of social science, robotics, and artificial intelligence theories into the LLMs test framework is a task riddled with complexities. This integration is a novel approach, and it brings forth the challenge of developing AI models in a way that accurately reflects these theories' concepts. The translation of abstract theoretical concepts into practical AI programming can be riddled with difficulties, which only increase with the complexity and unpredictability of real-world factors.

Like all AI models, digital citizens introduced in this study also run the risk of incorporating and even amplifying biases present in their training environments or data. While steps are taken to identify and correct these biases, the task's difficulty and the potential impact of these biases should not be underestimated.

Determination of successful AI alignment is a complex feat. It hinges on defining what constitutes "desirable" behavior and ensuring that the AI models exhibit such behavior consistently in a range of scenarios. Currently, our understanding and definition of successful AI alignment is constrained by known human values and societal norms, introducing a probable limitation of oversight or omission of novel or altered values and norms that can form in future societies.

*Future research*

Future research can build upon the findings of this study in several ways. Advancements in virtual and augmented reality technologies could substantially enhance the digital city's realism in this study, thereby improving the accuracy and relevance of the simulations.

Exploratory endeavors can focus on refining the process of integrating these varied theories into AI programming. New techniques or algorithms might be developed to facilitate this integration, based on learnings gained from the successes and challenges of the current approach.

Systematic efforts could be carried out to identify, understand, and correct biases in AI models with the help of insights gained from the digital city simulations. These techniques could then be incorporated into an automated bias-detection and correction framework for AI development.

There is a pressing need for a comprehensive, universally accepted definition and understanding of successful AI alignment, which respects the diversity and dynamism of human values across cultures and time. Future research should also focus on developing continuous monitoring and recalibration tools for AI models, ensuring their behavior remains aligned with these values over time, despite the changes and refinements in societal norms and regulations.

This research provides a fertile ground for advancements in the field of AI development and testing. Its pioneering blend of theories and simulation-based approach offers opportunities for further research and exploration towards the goal of secure, controlled, and value-aligned AI.

**Availability of Data and Materials**

The data that supports the findings of this study are available upon reasonable request from the corresponding authors. It includes a variety of theoretical and practical resources rooted in multiple disciplines, including sociology, social psychology, computer science, physics, biology, and economics.

**Competing Interests**

The authors, Ljubiša Bojić, Matteo Cinelli, Dubravko Ćulibrk, and Boris Delibašić, declare that they have no competing interests. They affirm that the submitted work is original, and it has not been influenced by any personal, financial, or commercial relationships or interests.

**Funding**

This research did not receive any specific grant from funding agencies in the public, commercial, or not-for-profit sectors. It is an independent study conducted by the authors listed, contributing to their respective institutional and academic portfolios.

**Authors' Contributions**

Ljubiša Bojić, Ph.D., conceived the original idea, provided overall guidance and wrote significant portions of the text. Matteo Cinelli, Ph.D., contributed to the design of the theoretical framework and wrote relevant parts of the text. Dubravko Ćulibrk, Ph.D., provided essential insights into the potential applications of AI and contributed to writing the paper. Boris Delibašić, Ph.D., helped refine the concepts and contributed to drafting the manuscript. All authors contributed to finalizing the paper, ensuring rigor in the methodology, and reviewing the final draft before submission.

**Acknowledgements**

Funded by the European Union. Views and opinions expressed are however those of the author(s) only and do not necessarily reflect those of the European Union or the European Health and Digital Executive Agency (HaDEA). Neither the European Union nor the granting authority can be held responsible for them. Grant Agreement no. 101120763 - TANGO.

This paper was realized with the support of the Ministry of Science, Technological Development, and Innovation of the Republic of Serbia, according to the Agreement on the Realization and Financing of Scientific Research.

This paper has been supported by the TWON (project number 101095095), a research project funded by the European Union, under the Horizon Europe framework (HORIZON-CL2-2022-DEMOCRACY-01, topic 07). More details about the project can be found on its official website: https://www.twon-project.eu/.